\documentclass[%
 aps,pra,
 amsmath,amssymb,
reprint,%
10pt]{revtex4-2}
\usepackage{comment}
\usepackage{graphicx} 
\usepackage{bbm}
\usepackage{amsmath}
\usepackage{amssymb}
\usepackage{commath}
\usepackage{graphicx}

\begin{document}

\title{Intra- and intercycle analysis of intraband high-order harmonic generation}

\author{Asbjørn Tornøe Andersen}
 \affiliation{Department of Physics and Astronomy, Aarhus University, Ny Munkegade 120, DK-8000 Aarhus C, Denmark}%
\author{Simon Vendelbo Bylling Jensen}
\affiliation{Department of Physics and Astronomy, Aarhus University, Ny Munkegade 120, DK-8000 Aarhus C, Denmark}
\author{Lars Bojer Madsen}
\affiliation{Department of Physics and Astronomy, Aarhus University, Ny Munkegade 120, DK-8000 Aarhus C, Denmark}%

\date{\today}

\begin{abstract}
We study intraband high-order harmonic generation arising from a band-gap material driven by a linearly polarized laser field. We factorize the intraband high-order harmonic-generation signal into intracycle and intercycle terms. The intracycle term uniquely determines the spectral characteristics whereas the intercycle term merely modulates the spectral features by imposing energy conservation in the long-pulse limit. Through analysis of the intracycle interference, the cutoff is identified, and the origin of the harmonic selection rules is revealed. Further, it is outlined how different components of the bandstructure contribute to different regions of the harmonic spectrum, giving rise to non-trivial intensity scaling of individual harmonics in the plateau region.
\end{abstract}
\maketitle

\section{Introduction}
Targeting matter with an intense laser pulse induces the nonlinear process of high-order harmonic generation (HHG) which can be applied to generate extremely short laser pulses \cite{attopulse_1, attopulse_2}. HHG was first observed in atoms and is most commonly understood through the three-step model \cite{threestep,PhysRevA.49.2117, HHGatomScaling}. In the three-step model, the atom is first ionized, the continuum electron then gains kinetic energy by propagating through the laser field, and finally, recombines with the ionized atom releasing high-energy radiation. The cutoff for HHG in atoms was observed to depend quadratically on the driving electric field strength. 

Observations of HHG with few-cycle, infrared driving pulses were extended to solid-state systems in 2011 \cite{bulk_hhg} and have since been a topic of active research. For band-gap materials, the process of HHG is typically understood through inter- and intraband electron dynamics \cite{inter+intra,hhgsolidtheory}. The interband contribution can be understood through a three-step model, similar to that of the atomic case. First, an electron transitions from the valence band to the conduction band creating an electron-hole pair, the electron-hole pair is then accelerated by the laser field, and finally, the electron recombines with the hole releasing high energy radiation.  Conversely, the intraband contribution is due to the acceleration of electron wavepackets within a band which, due to the non-parabolic bandstructure, results in the emission of high energy radiation. The intraband contribution dominates the emitted spectral regime below the band-gap energy whereas the interband contribution dominates above the band-gap energy. In both cases, the cutoff is observed to depend linearly on the driving electric field strength highlighting the difference between HHG in atoms and solids. It is interesting to study the inter- and intraband contributions to HHG in isolation. Here we wish to concentrate on the intraband contribution, which dominates the HHG process in the long wavelength regime when the electromagnetic field with frequency, $\omega_L$, is unlikely to cause interband transitions above the minimal band-gap energy, $\epsilon_{gap} \gg \hbar \omega_L$.  The intraband process of HHG in band-gap materials has proven particularly useful in outlining polarization dependencies \cite{PhysRevB.102.104308,PhysRevLett.120.243903} and extracting material properties such as conduction band reconstruction of ZnSe \cite{intraHHG_bandstrucZnSe} and measuring the ratio between harmonic dispersion components of SiO$_2$ \cite{Luu2015}. However, the Bloch electron model of the intraband electron dynamics has so far not provided a clear differentiation of spectral plateau and cutoff \cite{wannier_rep}.

For strong-field processes, it is often useful to employ an analysis based on intra- and intercycle dynamics. Here intracycle contributions occur within a single cycle of the driving electric field and intercycle contributions arise from electron processes across multiple cycles of the electric field. An intra- and intercycle analysis has been particularly useful to study above-threshold ionization (ATI) \cite{Arb__2010}. The intrinsic periodic properties of the ATI process allows the ATI spectrum to be factorized into a product of intercycle and intracycle interferences. Such analysis was first conducted within the electric dipole and strong-field approximation where intracycle interference was shown to act as a modulator of the multiphoton peaks generated by intercycle interference. This picture was subsequently shown to hold when going beyond the strong-field approximation as the Coulomb potential merely causes an intracycle interference phase shift \cite{intra_er_coulomb}. These results have since been generalized to the full momentum distribution \cite{doubly_diff} and been applied to identify intracycle trajectories that account for holographic interference \cite{holo_arbo_1, holo_andy_1, holo_andy_2, holo_andy_3}. Moreover, this framework of analysis was extended to a consideration of nondipole-induced effects in the ATI spectra \cite{lars_bojer_dis_re_ati}, two-color atomic ionization \cite{two_color_1, two_color_2} as well as laser-assisted photoionization both within \cite{LAPE_dip_1, LAPE_dip_2, LAPE_dip_3} and beyond the electric dipole approximation \cite{LAPE_nondip}. Experimental procedures have been developed to extract inter- and intracycle interferences from laser-assisted photoionization of argon \cite{LAPE_intra_inter_exp}.

An intra- and intercycle analysis of HHG in atoms or molecules can be performed in the Floquet limit \cite{atoms_intese_laser_fields} or from saddle-point analysis \cite{PhysRevA.49.2117}. It is natural to ask whether such analysis can be performed for the case of HHG in solids, and this question is  addressed for the intraband mechanism in the present work. We show that intraband HHG can be factorized in terms of intra- and intercycle interferences, analyze the characteristic features of the two terms and link the nontrivial properties of the harmonic spectra to the intracycle term. 

The paper is organized as follows. In Sec.~II, we present the theoretical model. The results are discussed in Sec.~III, followed by a conclusion in Sec.~IV. Atomic units are used throughout unless indicated otherwise.

\section{THEORETICAL MODEL}
\begin{figure}
    \centering
    \includegraphics[width= 1.00\columnwidth]{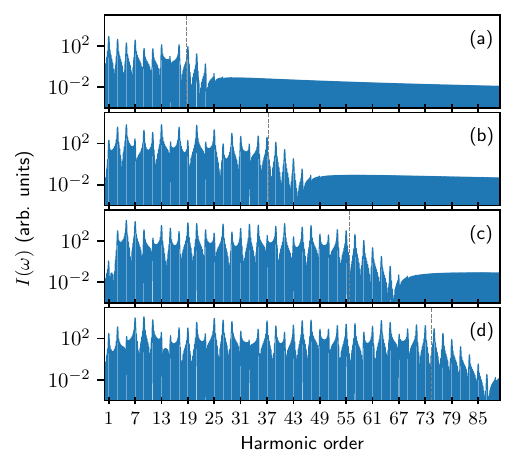}
     \caption{HHG spectra generated by a $N_c = 5$ - cycle laser pulse of peak electric field strength of (a) $F_0 = 0.01$, (b) $F_0 = 0.02$, (c) $F_0 = 0.03$, and (d) $F_0 = 0.04$. The predicted cutoffs in units of harmonic order [Eq.~\eqref{cutoffformula}] are illustrated by vertical dashed gray lines and given by $\gamma = 18.6$, $\gamma = 37.2$, $\gamma = 55.8$, and $\gamma = 74.5$ in (a)-(d), respectively.}
    \label{fig:bloch_ele}
\end{figure}

Throughout this paper, the interaction between a linearly polarized laser field with a band-gap material is studied in a one-dimensional setting using the electric dipole approximation and neglecting interband processes. The linearly polarized $N_c$-cycle laser pulse is described by the vector potential $A(t) = \frac{F_0}{\omega_L} \sin(\omega_L t)$ where $F_0$ is the peak electric field strength and $\omega_L$ the driving laser frequency. The electric field, $F(t)$, is related to the vector potential by $F(t) = -\partial_t A(t)$. Such a laser field corresponds to a flat-top pulse if neglecting ramp-on and ramp-off, as in Ref.~\cite{Arb__2010}. We consider the dynamics of an intraband electron wavepacket of Bloch states, which is likely to be generated in the conduction band at the earliest electric field maxima, which occur at time $t=0$ for the chosen laser parameters. Here the electron wavepacket is generated centered at crystal momentum $k(t=0) = k_0$. The ensuing dynamics of the electron wavepacket obey the acceleration theorem \cite{Yakovlev2016,Bloch1929,PhysRev.57.184}, 
\begin{equation} \label{crystal}
\dot{k} = -F(t) ,
\end{equation}
and the group velocity of the wavepacket, $v_g$, is determined by \cite{ash_mer}
\begin{equation} \label{group}
v_g =  \left. \frac{\partial{\varepsilon(k)}}{\partial k} \right\rvert_{k(t)},
\end{equation}
where $\varepsilon(k)$ is the band dispersion. Throughout this work, a material of inversion symmetry is considered and thus the dispersion can be expanded in a Fourier series
\begin{equation} \label{band}
\varepsilon(k)  =  \sum_{n=0}^{n_{max}} c_{n} \cos(n k a), 
\end{equation}
where $a$ is the lattice constant and $c_n$ are coefficients that include all material-specific properties, and correspond to the $n^{th}$ harmonic component of the effective band dispersion. 
Thus the generated intraband current is given as
\begin{equation} \label{current1}
\mathcal{J}(t) = -v_g(k(t)) ,
\end{equation}
where $k(t) = k_0 + A(t) $ follows from integration of Eq.~\eqref{crystal}. The generated intraband current is related to the emitted HHG spectrum by \cite{Yue:22}
\begin{equation} \label{spectrum}
    I(\omega) \propto  \left|\mathcal{F}\left( \frac{\partial \mathcal{J}}{\partial t} \right) \right|^2 ,
\end{equation}
where $\mathcal{F}$ denotes the Fourier transform given as 
\begin{equation}
    \mathcal{F}\left(\frac{\partial \mathcal{J}}{\partial t}\right) = \int_{-\infty}^{\infty}e^{-i \omega t } \frac{\partial \mathcal{J}}{\partial t} dt  .
\end{equation}
The HHG spectra are obtained for simulations of four different values of peak field strength and shown in Figs.~\ref{fig:bloch_ele}~(a)-(d), where the spectra consist of a plateau region of odd-ordered harmonics before the cutoff. Simulations throughout this paper consider the initial wavepacket crystal momentum to be at the minimum band-gap energy, corresponding to $k_0 = 0$, which is the most probable point of generation. Furthermore, the simulations are performed with driving laser frequency $\omega_L = 0.0227$  and a band-gap material similar to Ref.~\cite{uncovering} with band coefficients, 
$c_0 =  0.0089$, $c_1 = -0.3333$, $c_2 = 0.0260$, $c_3 = -0.0326$, $c_4 = 0.0086$, $c_5 = -0.0098$, $c_6 = 0.0052$, $c_7 = -0.0037$, $c_8 = 0.0033$, band-gap energy of $\epsilon_{gap } = 0.121$ and lattice constant of $a=5.3$. The dashed gray vertical lines in Fig. \ref{fig:bloch_ele} show the analytical cutoff, which arise from the following intra- and intercycle analysis and is given by Eq.~\eqref{cutoffformula} below. The main purpose of Fig. \ref{fig:bloch_ele}  is to remind the reader of typical intraband HHG spectra and their observed linear scaling  of the cutoff with electric field strength. In the rest of the paper, such spectra will be analyzed in terms of intra- and interband constributions. 

\section{RESULTS AND DISCUSSION} \label{sec: er_ra}
Inserting Eq.~\eqref{crystal}~-~\eqref{band} into Eq.~\eqref{current1} and exploiting the periodicity of the laser pulse, $A(t) = A(t+\frac{2 \pi}{\omega_L})$, one can factorize the emitted HHG spectrum
\begin{equation} \label{amplitude}
\begin{gathered}
I(\omega) \propto \left| A^{er}(\omega) \right|^2  \left|A^{ra}(\omega)\right|^2,
\end{gathered}
\end{equation}
into its intercycle amplitude
\begin{equation} \label{intercycle}
    A^{er}(\omega) = \sum_{m=0}^{N_c - 1} e^{-i m 2 \pi \frac{\omega}{\omega_L}} ,   
\end{equation}
and intracycle amplitude 
\begin{equation} \label{intra_1}
    A^{ra}(\omega) = \int_{0}^{\frac{2 \pi}{\omega_L}} e^{-i \omega t} \frac{\partial \mathcal{J}}{\partial t} dt.
\end{equation}
\begin{figure}
    \centering
    \includegraphics[width= 1.0\columnwidth]{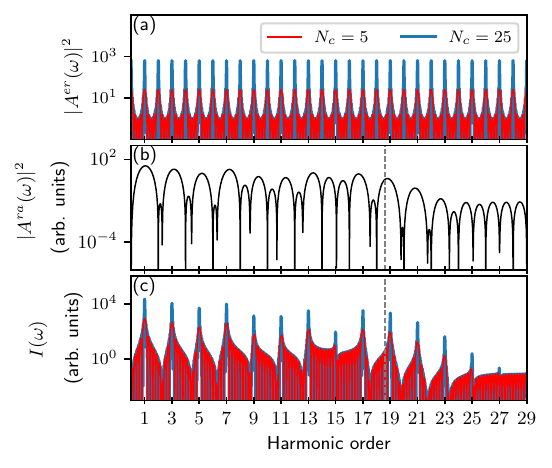}
    \caption{(a) The norm squared intercycle amplitude for $N_c = 25$ (blue line) and $N_c = 5$ - cycle pulse (red line). (b) The norm squared of the intracycle amplitude along with the predicted cutoff as illustrated by the vertical dashed line. The cutoff in units of harmonic order is at $\gamma = 18.6$ as obtained by Eq.~\eqref{cutoffformula}. (c) HHG spectra of Eq.~\eqref{amplitude} for the short and long-pulse of (a). The electric field strength is set to $F_0 = 0.01$ for (a)-(c).}
    \label{fig:intercyc}
\end{figure}
\\
The intercycle term is compared for a few-cycle $N_c = 5$ driving field pulse and a long $N_c = 25$ pulse in Fig.~\ref{fig:intercyc}~(a). With increasing number of cycles, the peaks of the intercycle interference, at integer multiples of $\omega_L$, become narrower and steeper. The intercycle term converges to a Dirac comb for $N_c \rightarrow \infty$ \cite{half_pois}. Thus the intercycle interference is responsible for energy conservation in the long-pulse limit where energy must be exchanged as multiples of $\omega_L$. Since the intracycle amplitude as defined in Eq.~\eqref{intra_1}, is independent of $N_c$ and since the intercycle amplitude of Eq.~\eqref{intercycle} is independent of material-specific parameters and acts similarly across all spectral regimes, the role of the intercycle interference is clear. It simply modulates the intracycle interference by imposing energy-conservation at increasing number of laser cycles. In other words, the structure of the HHG spectrum, such as the harmonic cutoff, selection rules, and material specific characteristics, derives solely from intracycle interference. This is demonstrated by comparing the HHG spectrum for $N_c = 5$ and $N_c = 25$ in Fig.~\ref{fig:intercyc}~(c). Here all spectral characteristics arise from the corresponding intracycle interference of Fig.~\ref{fig:intercyc}~(b), which is modulated by the intercycle interference of Fig.~\ref{fig:intercyc}~(a).
\begin{figure}
    \centering
    \includegraphics[width= .95\columnwidth]{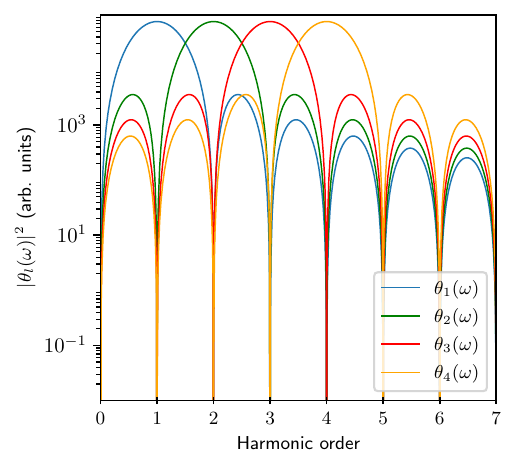}
    \caption{The intracycle term of $\theta_l(\omega)$ of Eq.~\eqref{theta} for $l=1$ (blue line), $l=2$ (green line), $l=3$ (red line) and $l=4$ (orange line).}
    \label{fig:1_intra}
\end{figure}

Considering now only the intracycle term, it can be further decomposed as
\begin{equation} \label{intracyc}
A^{ra}(\omega) = \sum_{l=-\infty}^{\infty} \theta_l(\omega) \Phi_l ,
\end{equation}
where 
\begin{equation} \label{theta}
    \theta_l(\omega) = \frac{e^{-i 2 \pi \frac{\omega}{\omega_L}} - 1 }{i( \omega_L l - \omega)} \mathbbm{1}_{\omega \neq l \omega_L} +\frac{2 \pi}{\omega_L} \mathbbm{1}_{\omega = l \omega_L},
\end{equation}
in which $\mathbbm{1}_{x \neq \alpha} = 1 \text{ if } x \neq \alpha \text{ and }0 \text{ if } x = \alpha$. Likewise, $\mathbbm{1}_{x = \alpha} = 1 \text{ if } x = \alpha \text{ and }0 \text{ if } x \neq \alpha$. The second factor in Eq.~\eqref{intracyc} is
\begin{equation}
\begin{gathered} \label{phi}
    \Phi_l = \frac{a}{2} \sum_{n=1}^{n_{max}} c_n n J_l\left(\frac{naF_0}{\omega_L}\right)l \omega_L  \\ \times \left[e^{-inak_0}(-1)^{l}-e^{inak_0} \right],
\end{gathered}
\end{equation} 
where $J_l$ is the Bessel function of the first kind of $l^{\text{th}}$ order, which originates from the use of the Jacobi-Anger expansion. To initially understand the behaviour of $\theta_l(\omega)$, it is illustrated for $l=1, 2, 3~\&~4$ in Fig.~\ref{fig:1_intra}. It is evident from Eq.~\eqref{theta} that for all $l < 0$ the term $\theta_l(\omega)$ returns a value of zero for all positive integer multiples of $\omega_L$. In light of this, since we only consider $\omega \geq 0$, we only examine terms arising from $l \geq 0$. For all $ l \geq 0$ the term $\theta_l(\omega)$ returns a value of zero for all integer multiples of $\omega_L$ except $l \omega_L$ where it returns the value $2 \pi / \omega_L$. Hence the intensity of the $l^{\text{th}}$ harmonic is solely determined by the $l^{\text{th}}$ term in Eq.~\eqref{intracyc}.

\begin{figure}
\centering
    \includegraphics[width= 0.95\columnwidth]{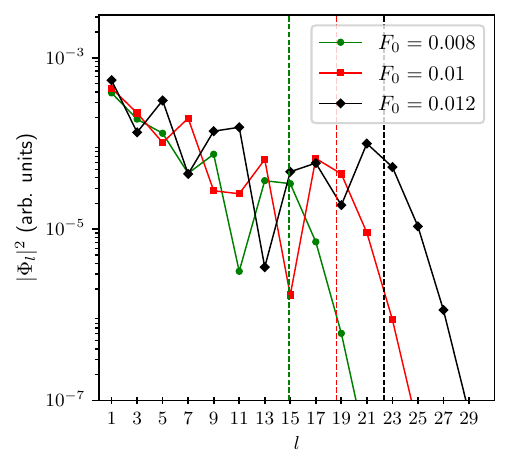}
    \caption{The intracycle term $\Phi_l$ of Eq.~\eqref{phi} for odd values of $l \geq 0$ and field strengths as given in the insert. The predicted cutoffs of Eq.~\eqref{cutoffformula} are at $\gamma=14.9$, 18.6 and 22.3, as indicated by the dashed vertical lines. } 
    \label{fig:2intra}
\end{figure}

Note that since $\Phi_l$ is independent of $\omega$, the pair $ \theta_l (\omega) \Phi_l $ can be interpreted as follows: For each $l \geq 0$ the term $\theta_l(\omega)$ generates a spectral peak centered at a harmonic order $l \omega_L$ and the term $\Phi_l$ generates the intensity for the peak at $l \omega_L$. This suggests that $\Phi_l$ is responsible for both the odd-harmonic selection rule, the harmonic cutoff and any material specific spectral features. 
More specifically, in Eq.~\eqref{phi} one can explicitly identify the selection rule for the allowed harmonic orders to arise from the factor $\left[e^{-inak_0}(-1)^{l}-e^{inak_0}\right]$. For $k_0 = 0$ all band components $c_n$ provide odd harmonic selection rules regardless of their $n^{\text{th}}$ order.  We note that any realistic excitation of the symmetric material will lead to a symmetric population of negative and positive crystal-momentum values. The integration of the signal amplitude, i.e., the intraband current, over the full Brillouin zone following this initial wavpacket excitation will also only lead to odd harmonics, as it should. Any even harmonics generated from an initial nonvanishing $k_0$-value in the wavepacket cancels with the contribution to the current from $-k_0$. The term $\Phi_l$ is shown for odd harmonics in Fig.~\ref{fig:2intra} for $k_0= 0$ where the harmonic cutoff is illustrated.
This cutoff in units of harmonic order, $l_\text{cutoff}=\gamma$, is directly explained  by the properties of the Bessel function (which decays when its order becomes larger then the magnitude of its argument)  and is given by 
\begin{equation} \label{cutoffformula}
    \gamma = \frac{n_{max} a F_0}{\omega_L}.
\end{equation}
\begin{figure}[]
    \centering
    \includegraphics[width= 0.95\columnwidth]{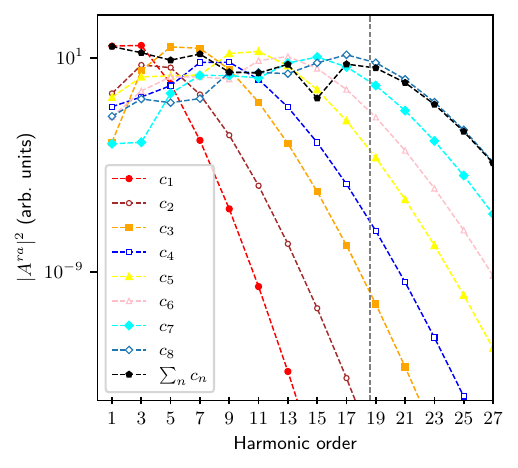}
    \caption{Norm-squared intracycle amplitude of Eq.~\eqref{intracyc} decomposed for different bandstructure coefficients $c_1$ (red line), $c_2$ (green line), $c_3$ (brown line), $c_4$ (orange line), $c_5$ (yellow line), $c_6$ (pink line), $c_7$ (light blue), $c_8$ (dark blue). The norm-squared intracycle amplitude for the full bandstructure is given for comparison in black.  The predicted cutoff of Eq.~\eqref{cutoffformula} at $\gamma= 18.6$ is illustrated by the dashed vertical line. The electric field strength is $F_0=0.01$.}
    \label{fig:cut_band}
\end{figure}
A similar equation for the cutoff was found in Refs.~\cite{bulk_hhg, Luu2015}. Our analysis, however, identifies that the cutoff arises solely from the intracycle interference and finds it to be independent of the initial crystal momentum, $k_0$. Morever, Eq.~\eqref{cutoffformula} can be viewed as a momentum space analog to the cutoff found using a Wannier representation in Ref.~\cite{wannier_rep}.
The intracycle analysis thus implies a linear relationship between the cutoff and the peak electric field strength, consistent with experimental observations~\cite{bulk_hhg, Luu2015}. It also implies that, by measuring a cutoff experimentally, one can determine the, $n_{max}$, i.e., the largest harmonic bandstructure component needed to accurately describe the bandstructure of Eq.~\eqref{band}.

To further highlight the origin of the cutoff it is meaningful to deconstruct the intracycle interference in terms of the contributions from each bandstructure coefficient, $c_n$ as done in Ref.~\cite{Luu2015}. To this end, in Fig.~\ref{fig:cut_band}, the intracycle contribution is provided from each coefficient of the bandstructure and compared to the total intracycle signal. In doing so, we identify a convergence between the intracycle contribution of the bandstructure coefficient with the highest $n$ and the intracycle contribution of the full bandstructure in the high-frequency spectral regime. That is, the cutoff for the HHG spectrum is alone determined by the last coefficient in the bandstructure as given in Eq.~\eqref{cutoffformula}. 

Another observation from Fig.~\ref{fig:cut_band} is the fact that each coefficient of the bandstructure only contributes significantly within a narrow spectral range. This is explained by the properties of the Bessel functions where each band coefficient, $c_n (n=1,...,n_{max})$, is accompanied by a Bessel function, $J_l(n a F_0 / \omega_L)$, with an associated cutoff $\gamma_n = n a F_0 / \omega_L$. Therefore the band coefficient, $c_n$, only contributes significantly up to harmonic orders of $\gamma_n$. Since typical bandstructures exhibit decreasing magnitude of $c_n$ for increasing $n$, there will, in the limit of large field strength, exist separate regions of the harmonic spectrum where different coefficients dominate the optical response. We note that identification of such separate spectral regions would allow for experimental reconstruction of the bandstructure, through reconstruction of its $c_n$-components. 

Returning to Fig.~\ref{fig:2intra}, we notice how the signal for many harmonics within the plateau show nonmonotonic behavior with increasing field strength. We see how, e.g., harmonic order 13 attains its smallest signal for the highest field strength. This effect is due to a combination of the oscillating behavior of the Bessel function of order $l$, once away from the cutoff regime, and the intracycle interference arising from the sum over the bandstructure components in Eq.~\eqref{phi}. Regions of interference occur between different components of the bandstructure in the intracycle interference and these regions are expected to shift linearly with the peak field strength alike the harmonic cutoff. The regions of interference thus also open an avenue for bandstructure reconstruction, by carrying information about the relative importance of adjacent bandstructure components. For example, destructive interference between the harmonic amplitudes related to different $c_n$-contributions to the bandstructure of Eq.~\eqref{band} is observed in Fig.~\ref{fig:cut_band} for harmonic order 15. For this harmonic we see how the norm-squared of the intraband amplitudes corresponding individually to $c_5, c_6, c_7$ and $c_8$  are all larger than the norm-square of the amplitude of the sum over contributions from all $c_n'$s. We note that  interference for the yield of the $l^{th}$ harmonic only occur within the plateau region $l < \gamma$ where the harmonic yield oscillates in magnitude with varying peak field strength. For the cutoff region of $l > \gamma$, the harmonic yield scales as $\left(F_0 \right) ^{2l}$ as given by the asymptotic form of the Bessel function and also predicted from Ref.~\cite{wannier_rep}.  Note also that we recover the expected scaling of harmonic signal for a given harmonic order in the low-field limit. Namely, for sufficiently low field strengths, the cutoff order will be low and the $\left(F_0 \right) ^{2l}$ scaling emerges for all harmonic order $l$.  In this manner, we recover the monotonous increase in the harmonic yield with field strength as predicted by Ref.~\cite{inten_dependence}.

\section{Conclusion}
In conclusion, we have factorized the intraband HHG spectrum into contributions of intra- and intercycle interferences showing how the spectral characteristics of the intraband HHG spectrum derives solely from the intracycle interference and that intercycle interference merely modulates the intracycle interference by imposing energy conservation in the long-pulse limit. Furthermore, the analysis confirmed that the cutoff for the intraband HHG spectrum depends linearly on the electric field strength. The analysis showed that the cutoff links directly to the largest significant harmonic component of the bandstructure [Eq.~\eqref{band}]. In addition to this, it was found that different harmonic contributions of the dispersion contribute to different emitted spectral regions as determined by their accompanying Bessel function. In the limit of large field strengths, each coefficient typically dominates a separate spectral region leading to perspectives for bandstructure reconstruction. Finally, in the plateau region, a nonmonototenic behavior of the intensity scaling of individual harmonics was related to  the interference between contributions to the HHG amplitude associated with different bandstructure components.

\acknowledgments
This work was supported by the Independent Research Fund Denmark (GrantNo.9040-00001B). S.V.B.J. further acknowledges support from the Danish Ministry of Higher Education and Science. 
\bibliography{reflibrary}

\end{document}